\documentstyle[epsf,rotate]{article}
\begin{document}

\begin{center}

{\LARGE OLD ISOLATED NEUTRON STARS}\\

\vskip 0.2cm

{\bf SERGEI B. POPOV}\\
\vskip 0.2cm

{\bf Sternberg Astronomical Institute, Moscow}\\
\vskip 0.2cm

{\bf polar@xray.sai.msu.su polar@sai.msu.ru}
\vskip 0.2cm

{\Large Abstract}\\

\end{center}

 In this poster I briefly review several articles
on astrophysics of old isolated neutron stars,\\
which were published in 1994-99 by my co-authors and myself.

\begin{figure}[h]
\epsfxsize=\hsize
\centerline{{\epsfbox{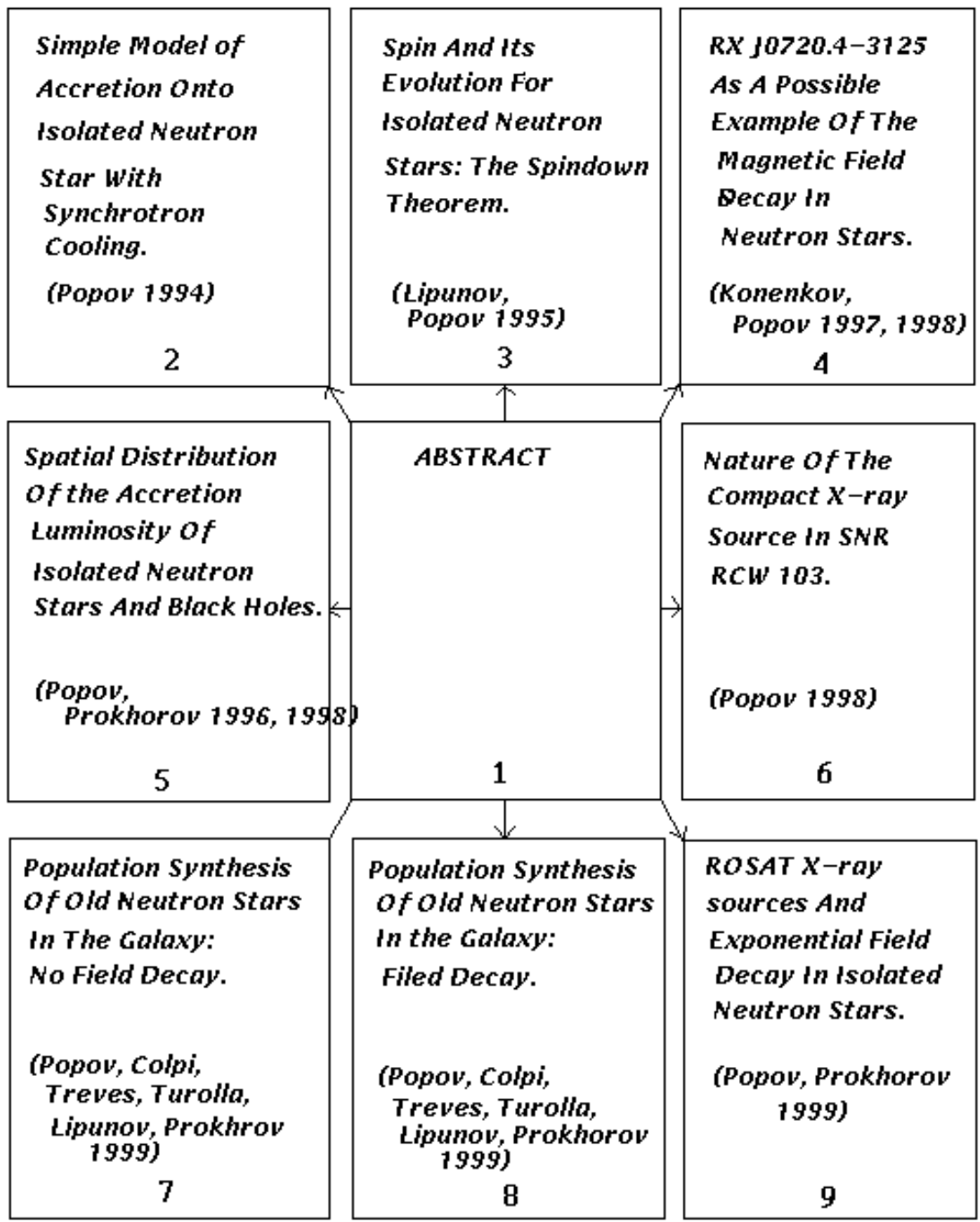}}}
\caption{Plan of the poster}
\end{figure}

{\bf Acknowledgments}

I want to thank all my co-authors: prof. V.M. Lipunov, dr. M.E. Prokhorov,
prof. M. Colpi, dr. D. Yu. Konenkov, prof. A. Treves, dr. R. Turolla.

\newpage

{\bf Simple Model Of  Accretion  Onto  Isolated  Neutron   Star  With
                       Synchrotron Cooling}\\

{\bf Popov S.B., Astron. Circ. N1556 pp. 1-2, 1994}\\

 Here I modeled accretion  onto  an  isolated
neutron star (INS) from the interstellar medium in the  case  of
spherical symmetry for different values of  the  magnetic  field
strength, ambient gas density and NS's mass. I  tried  to  verify
the idea that if the radius of corotation, $R_{co}$,
is less than the Alfven radius, $R_A$,
the shell will form around  the  INS  and   $R_A$
will decrease to $R_{co}$,   and the periodic X-ray source  will  appear
(see Treves et al., 1993   A\&A  269,   319).

Dependence of  $R_A$  on $t$ in our model roughly coincides  with  the
analytic formula from Treves et al. (1993): 
$R\sim t^{-1/2}$ (for some values $R_A$ was decreasing faster).

\begin{figure}[h]
\epsfxsize=0.6\hsize
\centerline{{\epsfbox{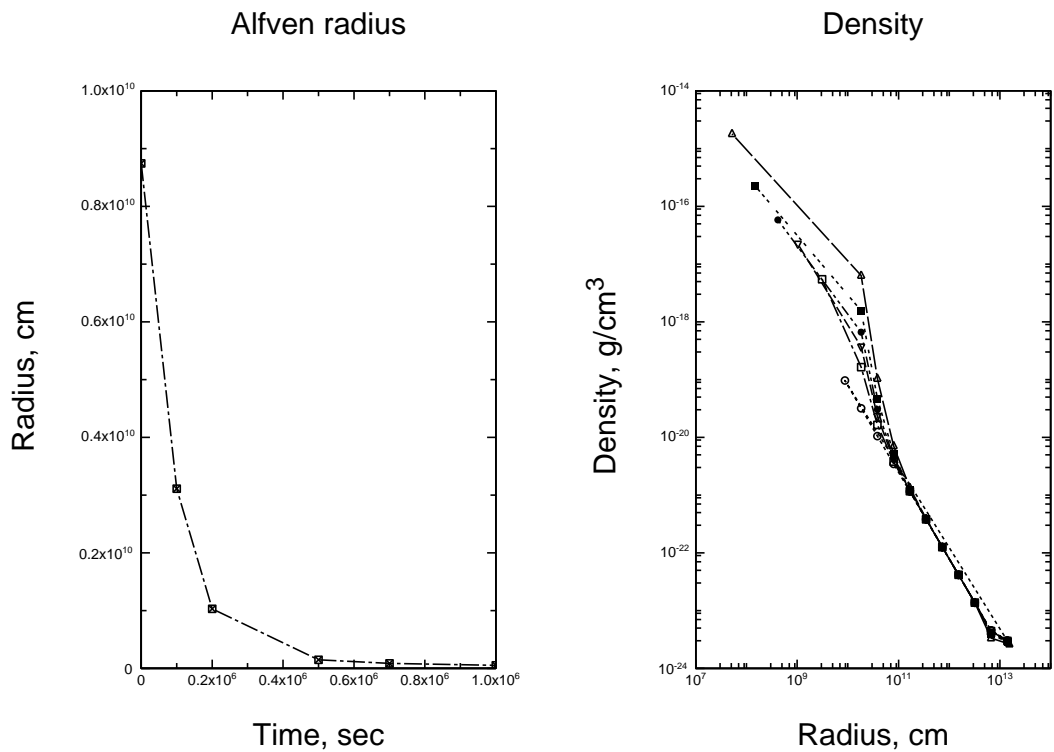}}}
\caption{Growth of density and decreasing of the Alfven radius}
\end{figure}

On the figure I show the growth of the envelope density (curves on the
figure are plotted for different moments of time: higher density corresponds
to later moments of time) and the
decreasing of the Alfven radius with time.

Periodic  sources  with  $P$  from
several minutes to several months can appear.

\newpage

{\bf Spin and its evolution for isolated neutron stars: the Spindown
theorem}\\

{\bf Lipunov V.M. \& Popov S.B., AZh  72, N5, pp. 711-716, 1995
(astro-ph/9609185)}\\

A possible scenario of spin evolution of isolated neutrons stars is
considered.

 The new points of our consideration are (all points,
including the Spindown theorem, are formulated for constant field!):

\noindent
--we give additional arguments for the relatively
short time scale of the Ejector stage ( $ \approx 10^7-10^8 $ yrs for small
velocities of NSs).

\noindent
--we propose specific SPINDOWN THEOREM and give some arguments for its
validity. This theorem argues, that the Propeller stage is always shorter than
the Ejector stage (for constant magnetic field).

\noindent
--we consider evolution of spin period of a NS on the Accretor stage
and predict that its period  without field decay is
$ \geq 5\cdot 10^2 $ sec and INSs can be observed
as pulsating X--ray sources.

\noindent
--we consider new idea of stochastic acceleration of  very old NSs due to
accretion of turbulizated ISM. A specific equilibrium period can be reached.

\noindent
-- accreting INSs can be spin-up and spin-down with equal probability.

\begin{figure}[h]
\epsfxsize=0.6\hsize
\centerline{{\epsfbox{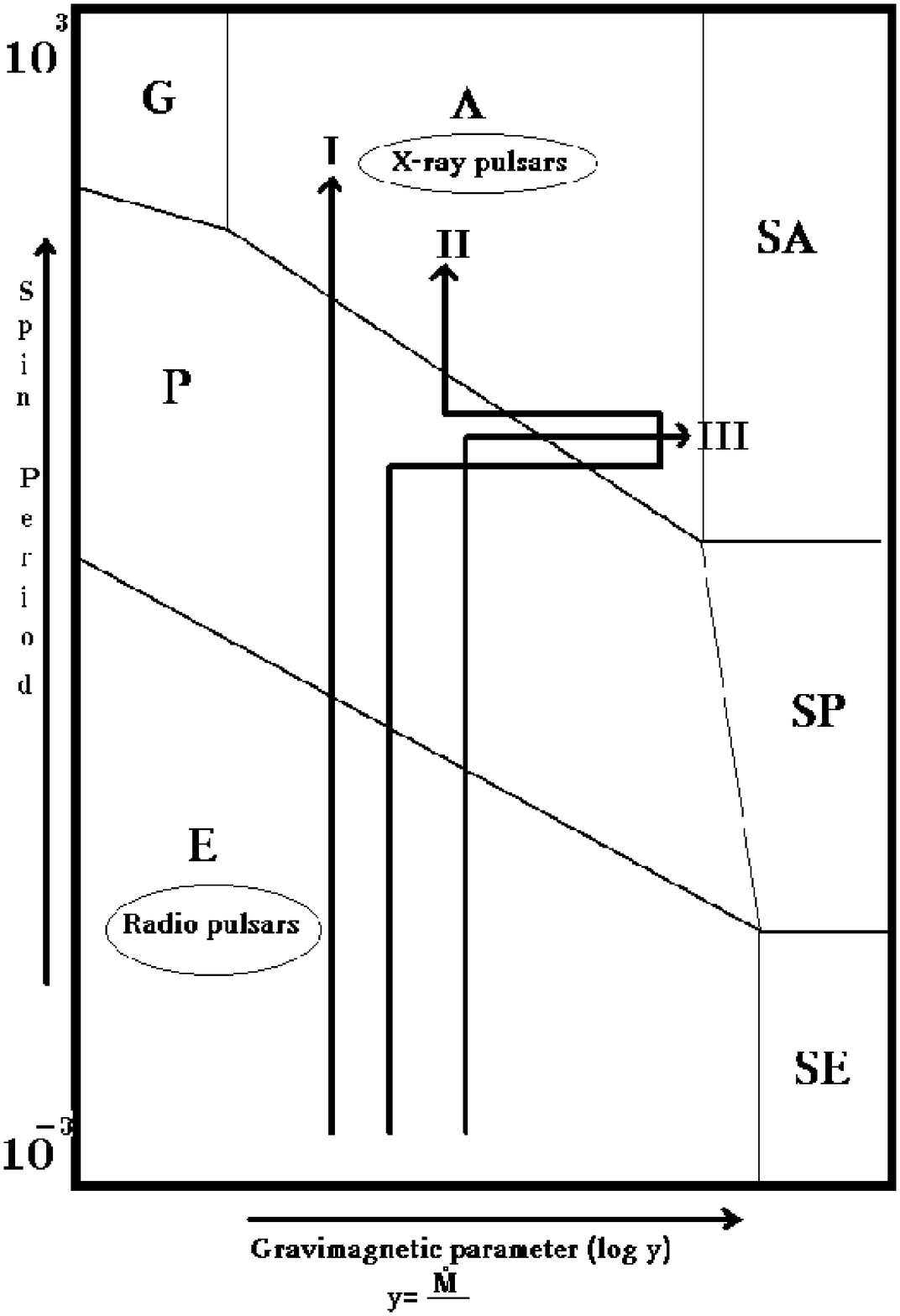}}}
\caption{$P-y$ diagram}
\end{figure}

The figure illustrates magneto-rotational evolution of an isolated neutron
star on $P-y$ diagram. Gravimagnetic parameter: $y=\frac{\dot M}{\mu ^2}$.

\newpage

{\bf RX J0720.4--3125 as a Possible Example of the  Magnetic
Field Decay of Neutron Stars}\\

{\bf Konenkov D.Yu. \& Popov S.B., PAZh  23, pp. 569-575, 1997
(astro-ph/9707318)}\\

{\bf Popov S.B. \& Konenkov D.Yu.,  Radiofizika 41, pp. 28-35, 1998
(astro-ph/9812482)}\\

We studied possible evolution of the spin period and the magnetic
field of the X-ray source RX J0720.4-3125 assuming this source to be an
isolated neutron star accreting interstellar medium. Magnetic field of the
source is estimated to be $10^6 - 10^9$ G, and it is difficult to explain
observed spin period  8.38 s without invoking hypothesis of the
magnetic field decay. We used the model of ohmic decay of the crustal
magnetic field. The estimates of accretion rate ($10^{-14} - 10^{-16}
M_{\odot}/{\rm yr}$), velocity of the source relative to interstellar
medium ($10 - 50 $ km/s), neutron star age
($2\cdot 10^9 - 10^{10}$ yrs) are obtained. 

We also make new estimate of the equilibrium period for accreting INS in the
ISM:

$$
P_{eq}=960 k_t^{1/3}\mu_{30}^{2/3}I_{45}^{1/3}\rho_{-24}^{-2/3}
v_{\infty_6}^{13/3}v_{t_6}^{-2/3} M_{1.4}^{-8/3}\,\, {\rm sec}
$$

The period $P_{eq}$ corresponds to the NS rms
rotation rate obtained from the solution of the corresponding
Fokker--Planck equation. In reality, the rotational period of INS
fluctuates around this value.
We take into account the three-dimensional character of turbulence,
i.e. the fact that the vortex can be oriented not only in the
equatorial plane but also at any angle to this plane. In this
case, diffusion occurs in the three-dimensional space of angular
velocities.

\begin{figure}[h]
\epsfxsize=0.5\hsize
\centerline{{\epsfbox{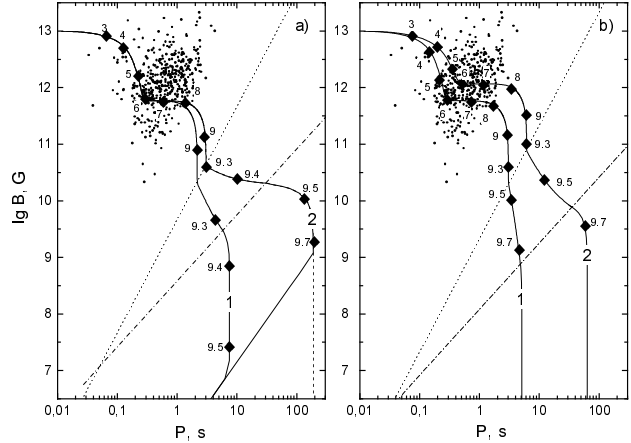}}}
\caption{ The evolutionary tracks of the neutron star for the accretion rates
$\dot M =
10^{-15} M_{\odot}\,$yr$^{-1}$ (a) and $\dot M = 10^{-16} M_{\odot}\,$
yr$^{-1}$ (b). The dashed lines correspond to $p =
P_E$; the dot-dashed lines, to $p = P_A$. The dashed line in the figure
shows for the second track the neutron star evolution with no acceleration in
the turbulized interstellar medium. The numbers near the marks in tracks
denote the logarithm of the neutron star age in years. The observed radio
pulsars are indicated by dots.}
\end{figure}

\newpage

{\bf Spatial distribution of the accretion luminosity of
isolated neutron stars and black holes in the Galaxy}\\

{\bf Popov S.B. \& Prokhorov M.E., A\&A 331, pp. 535-540, 1998
(astro-ph/9705236)}\\

{\bf Popov S.B. \& Prokhorov M.E., astro-ph/9606126, 1996}\\

    We present here a computer model of the
spatial distribution of the luminosity,
produced by old isolated neutron stars (NS) and black holes (BH)
accreting from the interstellar medium.

We solved numerically the system of differential equations of motions
in the  Galactic potential.
The density in our model is constant in time.
 In our model we assumed that the birthrate of NS and BH is
proportional to the square of the local density.
Stars were assumed to be
born in the Galactic plane (Z=0) with circular velocities
plus additional isotropic kick velocities.
Kick velocities were taken equal for NS and  BH.
It is possible however that BH have lower
kick velocities because of their higher masses.

We used masses $M_{NS}=1.4 M_{\odot}$ for NS and
$M_{BH}=10 M_{\odot}$
for BH. 
The radii,
$R_{lib}$, where the energy is liberated, was assumed to be equal to
10 km for NS and 90 km (i.e. $3 R_g$, $R_g=2GM/c^2$)
for BH.

For each star we computed the exact trajectory and the
accretion luminosity.
The accretion luminosity was calculated using Bondi's formula.
Calculations used a
grid with a cell size 100 pc in the R-direction and 10 pc in
the Z-direction.  

 As expected, BH give  higher luminosity than  NS,
as they are more massive.
But if the total number of BH is significantly lower than the number
of NS, their contribution
to the luminosity can be less than the contribution of NS.
The total accretion luminosity of the Galaxy for $N_{NS}=10^9$
and $N_{BH}=10^8$ is about $10^{39}-10^{40}${\it erg/s}.
For a characteristic velocity of 200 km/s  the maximum of the
distribution is situated approximately at  5.0 kpc for NS  and  at  4.8
kpc for BH.  For NS with a characteristic velocity of 400 km/s maximum is
located at 5.5 kpc, and for BH at 5.0 kpc.  This result is also
expected because of high masses of BH.

    The toroidal structure of the luminosity distribution
of NS and BH is an interesting and important feature
of the Galactic potential.
As one can expect, for low characteristic kick velocities and for BH
we have a higher luminosity.

    As we made very general assumptions, we argue, that such a distribution
is not unique  for our Galaxy, and all spiral galaxies can have such
a distribution of the accretion luminosity, associated with accreting NS
and BH.

\begin{figure}[h]
\epsfxsize=\hsize
\centerline{{\epsfbox{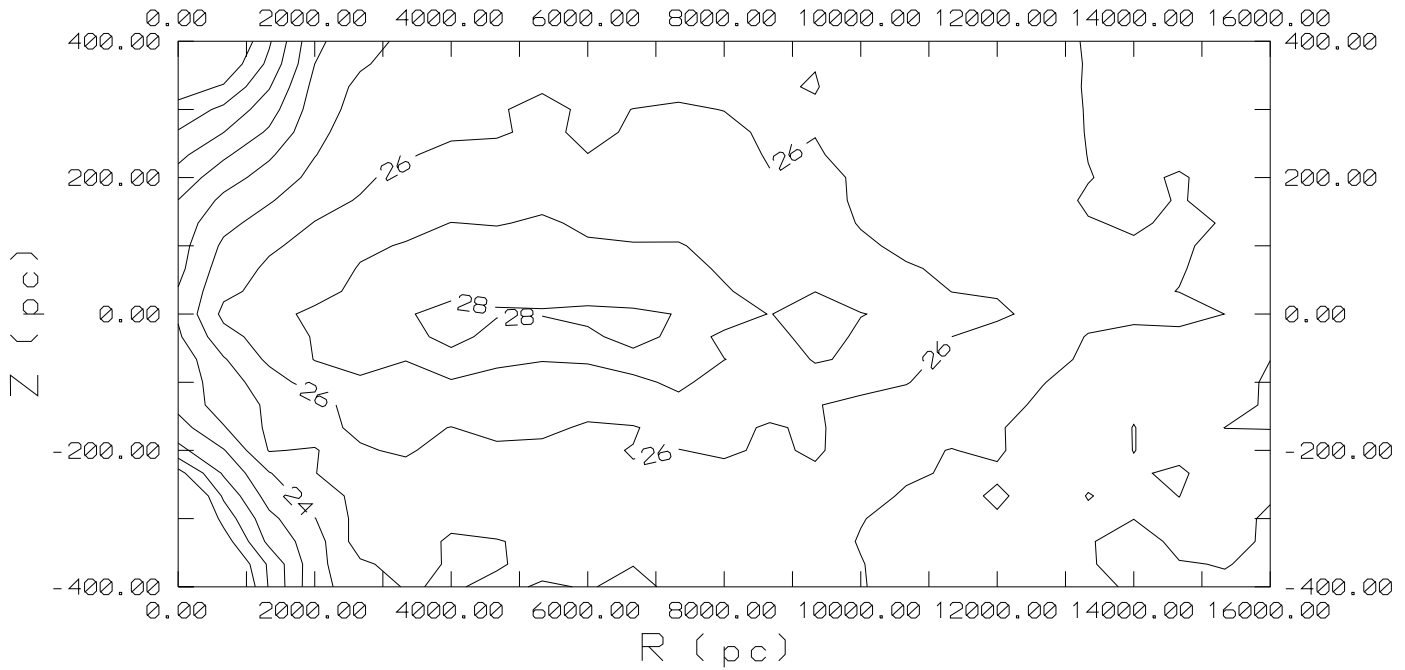}}}
\caption{The accretion luminosity distribution in the $R$--$Z$ plane for
neutron stars for a characteristic kick velocity 200 km/s.
The luminosity is  in ergs per second per cubic parsec.
$N_{NS}=10^9$.}
\end{figure}

\newpage

{\bf Nature of the compact X-ray source in supernova remnant RCW103}\\

{\bf Popov S.B., Astron. Astroph. Trans. 17, pp. 35-40, 1998
(astro-ph/9708044; astro-ph/9806354) }\\

Here I briefly discuss the nature of the compact X-ray source in the center of
the supernova remnant RCW 103. Several models, based on the accretion
onto a compact object such as a neutron star or a black hole
(isolated or binary), were analyzed.

I showed that it is more likely that the central X-ray source is an
accreting neutron star than an accreting black hole.
I also argue that models of a disrupted
binary system consisting of  an old accreting neutron star
and a new one observed as a 69-ms X-ray and radio pulsar are most
favored.

\begin{figure}[h]
\epsfxsize=0.7\hsize
\centerline{{\epsfbox{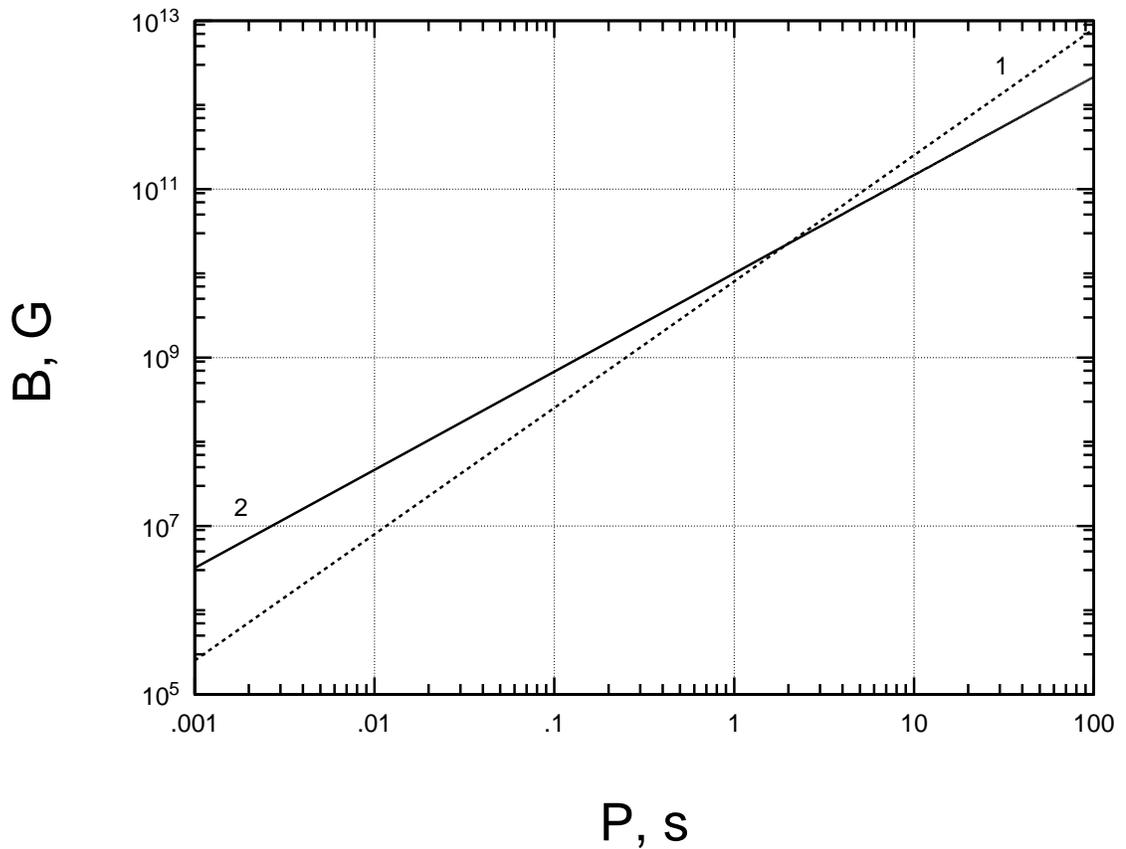}}}
\caption{Possible values of the magnetic field, B, and period, p, for
the accreting NS. The dotted line (1) corresponds to
the equilibrium period, $P_{eq}$, while
solid line (2) corresponds to the accretor period, $P_A$.}
\end{figure}

\newpage

{\bf Population synthesis of old neutron stars in the Galaxy: no field
decay}\\

{\bf Popov S.B., Colpi M., Treves A., Turolla R., Lipunov V.M., Prokhorov
M.E., ApJ 530 (20 Feb), 2000 (astro-ph/9910114)}\\

Isolated neutron stars (NSs) are expected to be as many as
$10^8$--$10^9$, $\sim 1\%$ of the   
total stellar content of the Galaxy.
Young NSs, active as pulsars, comprise only a tiny fraction
($\sim 10^{-3}-10^{-4}$) of the entire population, and about 1,000
have been  detected in
radio surveys. 

The paucity of  old isolated accreting neutron stars in
ROSAT observations is used to derive a lower limit on the mean
velocity of neutron stars at birth. The secular evolution of the
population is simulated following the paths of a statistical
sample of stars for different values of the initial kick velocity,
drawn from an isotropic Gaussian distribution with
mean  velocity
$0\leq \langle V\rangle\leq 550$ ${\rm km\,s^{-1}}$. The
spin--down, induced by dipole losses and the interaction with
the ambient medium, is tracked together with the dynamical
evolution in the Galactic potential, allowing for the
determination of the fraction of stars which are, at present, in
each of the four possible stages: Ejector, Propeller, Accretor,
and Georotator. Taking  from the ROSAT All Sky Survey an upper
limit of $\sim 10$ accreting neutron stars within $\sim 140$ pc from the
Sun, we infer a lower bound for the mean  kick velocity,
$ <V> \ge 200-300$ ${\rm km\,s^{-1}}$.

Present results,
moreover, constrain the fraction of low velocity stars, which
could have escaped pulsar statistics, to $\sim 1\%$.

\begin{figure}[h]
\epsfxsize=0.6\hsize
\centerline{\rotate[r]{\epsfbox{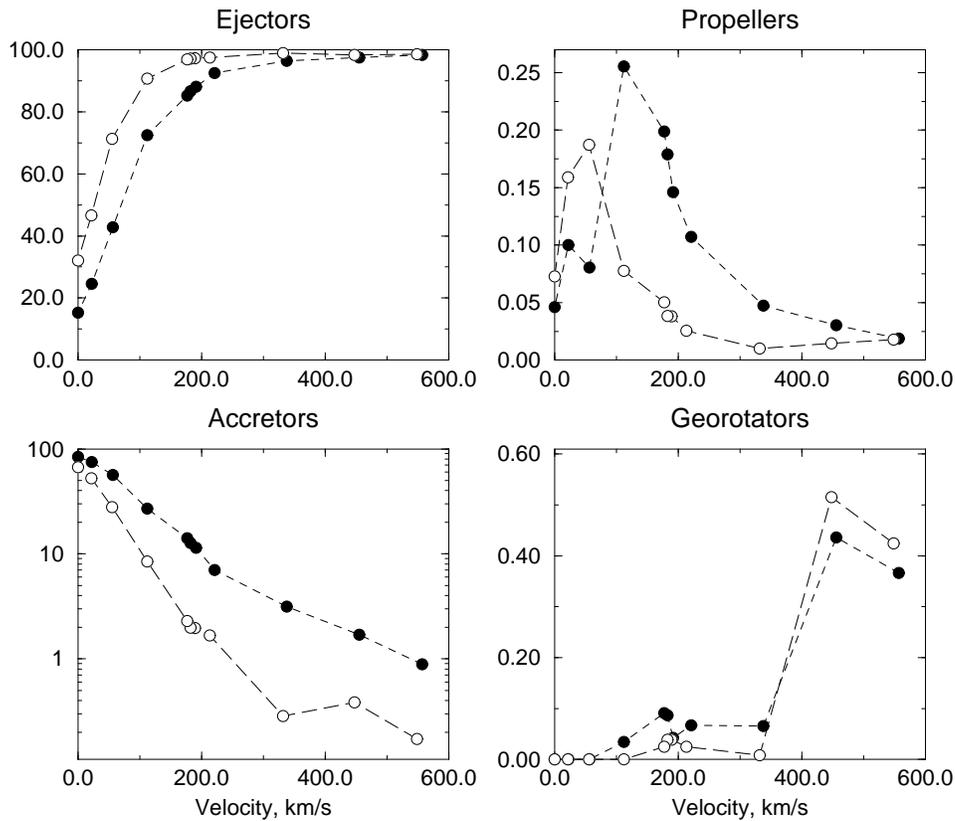}}}
\caption{Fractions of NSs in the different stages vs. the mean kick
velocity for $\mu_{30}=0.5$ (opaque circles) and $\mu_{30}=1$
(filled circles); typical statistical uncertainty for ejectors and
accretors is $\sim $ 1-2\%.}
\end{figure}

\newpage

{\bf Population synthesis of old neutron stars in the Galaxy: with field
decay}\\

{\bf Popov S.B., Colpi M., Treves A., Turolla R., Lipunov V.M., Prokhorov
M.E., ApJ 530 (Feb 20), 2000 (astro-ph/9910114)}\\

The time evolution of the magnetic field in isolated NSs is still
a very controversial issue and no firm conclusion has been
established as yet. A strong point is that radio pulsar
observations seem to rule out fast
decay with typical times less than $\approx 10$ Myr, but this does
not exclude the possibility that $B$ decays over much longer
timescales ($t_d \sim 10^9-10^{10}$ yr).

The same conclusion as at the previous page is reached for both a constant
($B\sim 10^{12}$ G) and a magnetic field decaying exponentially with a
timescale $\sim 10^9$ yr.

We refer here only
to a very simplified picture of the field decay
in which $B(t) = B(0)\exp{(-t/t_d)}$.
Calculations have been performed for $t_d=1.1\times 10^9$
yr, $t_d=2.2\times 10^9$ yr and $\mu_{30}(0) =1$. 
Results are shown in  the figure.

As it is
expected, the
number of Propellers is significantly increased with respect to   
the non--decaying case, while Ejectors are now less abundant.
Georotators are still very rare.
 The fraction  of Accretors
is approximately the same for the two values of $t_d$, and, at   
least for low mean velocities, is comparable to that of the
non--decaying field while,
at larger  speeds, it seems to be somehow higher.
This shows that the fraction of Accretors 
depends to some extent on how the magnetic field decays.       
By contrast, a  fast and progressive decay of $B$
would lead to an overabundance of Accretors because this situation is
similar to
``turning off'' the magnetic field, i.e., quenching any magnetospheric
effect on the infalling matter.

Summarizing, we can conclude that, although both the initial distribution
and the subsequent evolution of the
magnetic field strongly influences the NS census and should be accounted
for, the lower bound on the 
average kick derived from ROSAT surveys is not
very sensitive to $B$, at least for not too extreme values of $t_d$ and
$\mu(0)$, within this model.

\begin{figure}[h]
\epsfxsize=0.6\hsize
\centerline{\rotate[r]{\epsfbox{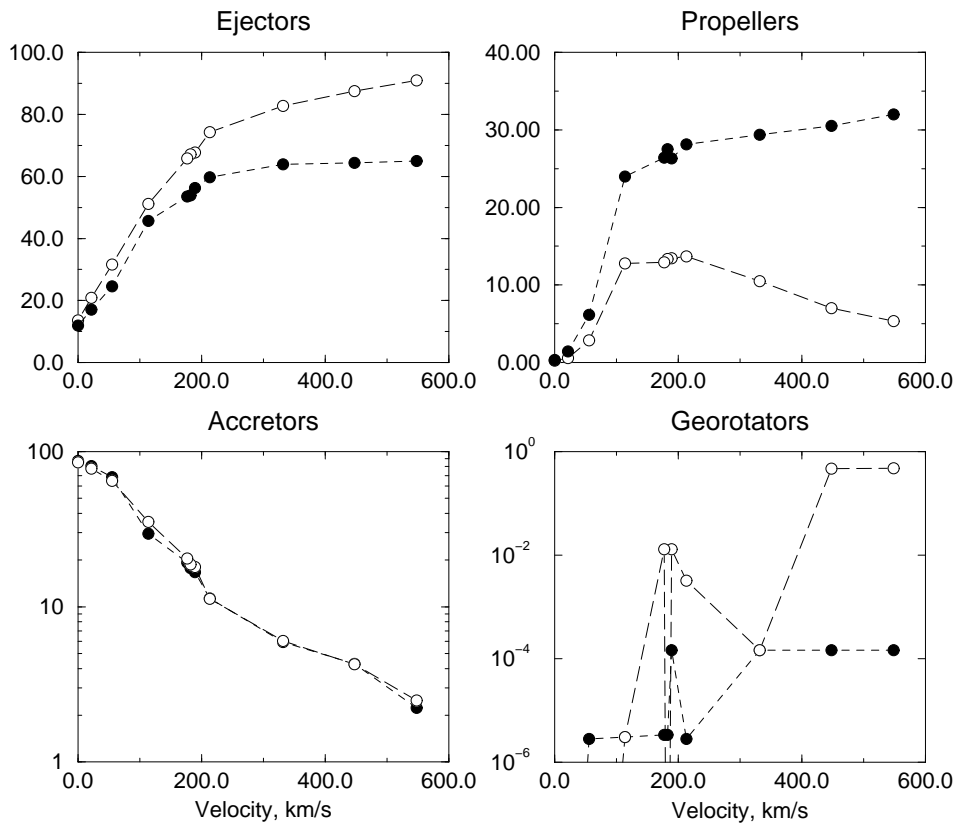}}}
\caption{Fractions of NSs in the different stages vs. the average kick
velocity for a decaying field with an e--folding time
$t_d=2.2\times 10^9$ yrs (opaque circles) and $t_d=1.1\times 10^9$ yrs
(filled circles).}
\end{figure}

\newpage

{\bf ROSAT X-ray sources and
exponential field decay in isolated neutron stars}\\

{\bf Popov S.B. \& Prokhorov M.E., astro-ph/9908212, 1999}\\

Many astrophysical manifestations of neutron stars (NSs) are determined by
their periods and magnetic fields.
Magnetic field decay in NSs is a matter of controversy. Many models of the
magnetric field decay have
been proposed.  

The influence of 
exponential magnetic field decay on the spin evolution of 
isolated neutron stars is studied. 
The ROSAT observations of several X-ray sources, which can be
accreting old isolated neutron stars, are used to 
constrain the exponential decay parameters.
Even if all modern candidates are not accreting objects,
the possibility of limitations of magnetic field
decay models based on future observations of isolated accreting NSs
should be addressed. 

We show that    
the range of minimum  value of magnetic moment, $\mu_b$,
and the characteristic  decay time, $t_d$, 
$\sim 10^{29.5}\ge \mu_b \ge 10^{28} \, {\rm G}\, {\rm cm}^3$ ,
$\sim 10^8\ge  t_d \ge 10^7\, {\rm yrs}$ are excluded  
assuming the standard 
initial magnetic momentum, $\mu_0=10^{30} \, {\rm G}\, {\rm cm}^3$. 
For these parameters,
neutron stars would never reach the stage of accretion from the interstellar
medium even for a low space velocity of the stars 
and a density of the ambient plasma. 
The range of excluded parameters increases for lower values of $\mu_0$.

In fact the limits obtained are even stronger than
they could be in nature, because we did not take into account
that NSs can spend 
some significant time (in the case with field decay)
at the propeller stage.

We conclude that 
the existence of several old isolated accreting NSs observed by ROSAT
(if it is the correct interpretation of observations),
can put important bounds on the models of the magnetic field decay for
isolated NSs (without influence of accretion, which can stimulate field
decay).
These models should 
explain the fact of observations of $\sim 10$ accreting isolated NSs in the
solar vicinity. Here we can not fully discuss the relations between
decay parameters and X-ray observations of isolated NSs without detailed
calculations. What we showed is that this connection should be taken
into account and made some illustrations of it,
and future investigations in that field would be desireble.

\begin{figure}[h]
\epsfxsize=0.6\hsize
\centerline{\rotate[r]{\epsfbox{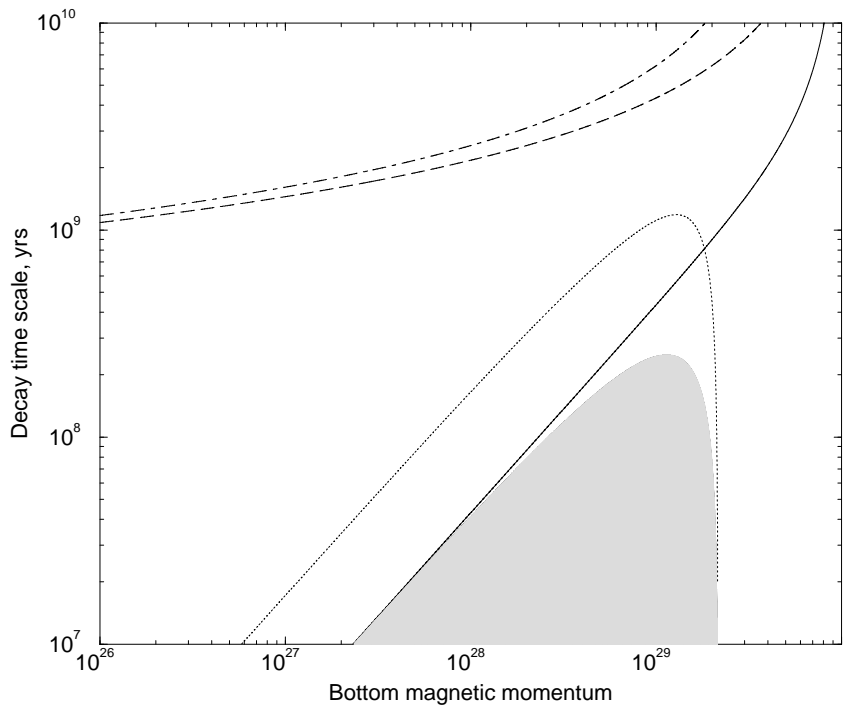}}}
\caption{Characteristic time scale of the magnetic field decay, $t_d$, vs.
bottom magnetic momentum, $\mu_b$.
In the filled region $t_E$ is greater than $10^{10} {\rm yrs}$.
Dashed line corresponds to $t_H=t_d\cdot \ln \left( \mu_0/\mu_b
\right)$, where $t_H=10^{10}$ years. Solid line corresponds to
$p_E(\mu_b)=p(t=t_{cr})$, where $t_{cr}=t_d\cdot \ln \left(
\mu_0/\mu_b \right)$. Both lines and filled region
are plotted for $\mu_0=10^{30} {\rm G} \, {\rm cm}^{-3}$.
Dot-dashed line is the same as the dashed one, but for $\mu_0=5\cdot 10^{29}
\, {\rm G} \, {\rm cm}^3$.
Dotted line is a border of the `forbidden'' region for $\mu_0=5\cdot
10^{29}  \, {\rm G} \, {\rm cm}^3$.}
\end{figure}

\end{document}